\documentclass[%
 aip,
 jmp,%
 amsmath,amssymb,
reprint,%
]{revtex4-1}

\usepackage{graphicx}
\usepackage{dcolumn}
\usepackage{bm}

\begin{document}

\preprint{AIP/123-QED}

\title[Sample title]{Unchanged thermopower enhancement at the semiconductor-metal \\ transition in correlated FeSb$_{2-x}$Te$_x$} 

\author{P. Sun}
\affiliation{ 
Max Planck Institute for Chemical Physics of Solids, D-01187 Dresden, Germany}%

\author{M. S\o ndergaard}
\author{Y. Sun}
\author{S. Johnsen}
\author{B. B. Iversen}
\affiliation{
Department of Chemistry, University of Aarhus, DK-8000 Aarhus C, Denmark}

\author{F. Steglich}
\affiliation{ 
Max Planck Institute for Chemical Physics of Solids, D-01187 Dresden, Germany}%

\date{\today}

\begin{abstract}
Substitution of Sb in FeSb$_2$ by less than 0.5\% of Te induces a transition from a correlated semiconductor to an unconventional metal with large effective charge carrier mass $m^*$. 
Spanning the entire range of the semiconductor-metal crossover, we observed an almost constant enhancement of the measured thermopower compared to that estimated by the classical theory of 
electron diffusion. Using the latter for a quantitative description one has to employ an enhancement factor of 10-30. 
Our observations point to the importance of electron-electron correlations in the thermal transport of FeSb$_2$, and suggest a route to design thermoelectric materials for cryogenic 
applications.    
\end{abstract}

                             
\maketitle

Current material science has not yet found efficient thermoelectric (TE) materials for cryogenic ($T$\,$<$100\,K) applications such as spot cooling microelectronic superconducting devices 
\cite{goodTE}.  In this context recent observations of a colossal thermopower $S$ of $-$(6$-$45) mV/K below 30 K in FeSb$_2$ have stimulated a great interest in the underlying  physics and 
also practical applications \cite{bentien, sun_prb, Kotliar}. Original interest in this compound stems from its narrow energy gap and strong electronic correlations as well as the thus 
induced unusual physical properties \cite{bentien2, petrovic1, petrovic2, LDA}. For a long time, systems with electron correlations have been expected to be candidates for TE applications 
\cite{goodTE}.  Following this line are correlated semiconductors, e.g., FeSi, and correlated metals, such as CePd$_3$, which do show promising TE properties at cryogenic range \cite{mahan}. 
Though the figure of merit $zT$ (= $T S^2$/$\rho \kappa$, with $\rho$ being the electrical resistivity and $\kappa$ the thermal conductivity) is only $\sim$0.01 at 50 K for FeSi, it can be 
enhanced to slightly below 0.1 at 100 K by 5\% Ir doping \cite{mahan}.  CePd$_3$, on the other hand, an intermediate-valence metal with low carrier density, exhibits a $zT$ = 0.23 at around 
200 K \cite{mahan}. Our intensive investigations on FeSb$_2$ have already clarified the interplay between the enhanced thermopower and electron correlations \cite{bentien, sun_prb, 
sun_dalton, sun_APEX}, nevertheless the microscopic mechanism of the enhancement is far from being understood \cite{Kotliar}.   

Here, we report TE properties of FeSb$_{2-x}$Te$_{x}$ with a narrow doping range, 0$<$$x$$<$0.16, where a semiconductor-metal (SM) transition occurs. Magnetism and electrical resistivity 
over a wide $x$ range (0$-$1.2) have already been reported by Hu $et\,al.$\cite{Hu_Te}. Our emphasis is aimed at learning to what extent the enhanced $S$ in a correlated semiconductor can be 
kept when crossing a SM transition. 
FeSb$_2$ and FeTe$_2$ are iso-structural (marcasite type) semiconductors having rather different energy gaps. The former shows a transport gap of $E_g$=26$-$36\,meV together with an even 
smaller one of $\sim$6\,meV \cite{sun_dalton}, contrasting to a much larger $E_g$ of the latter, 0.2$-$0.5\,eV \cite{harada}.  
The most striking finding of our study is that the enhancement to $S(T)$ (by a factor of 10$-$30) relative to the classical expectation, is rather robust against doping, spanning the entire 
range of the SM transition. This feature provides a convincing link between the enhanced $S(T)$ and the particular electronic structure of FeSb$_2$.  
Upon Te doping, a large decrease of  $\kappa$ is also observed which however, results in an only minor rise in $zT$, because of an overwhelming lattice contribution $\kappa _L$ even in the 
metallic regime.  

Single crystals of FeSb$_{2-x}$Te$_x$ were grown by chemical vapor transport using Br$_2$ as transport agent, and subsequently characterized and oriented using power x-ray and Laue 
diffraction methods \cite{sun_dalton}. It turns out that the actual concentration $x_a$ of Te as estimated by Hall-effect measurements may considerably deviate from the nominal value $x_n$ 
(inset of Fig. 1); even different crystals from the same batch show slightly differing values of $x_a$.  The reasons for this may involve the hard-to-control amount of the transport gas and 
different partial pressures of the formed bromides. 
The data presented here were obtained by employing a physical property measurement system (PPMS, Quantum Design), except for those of $x_a$ = 0.01, which has a smaller dimension ($<$ 4\,mm 
long) and was measured using a home-made cryostat equipped with a Au(Fe0.07\%)-Chromel thermocouple.
We paid particular attention to measure all the transport properties on exactly the same crystal for each doping concentration.  

\begin{figure}
\includegraphics[width=0.85\linewidth]{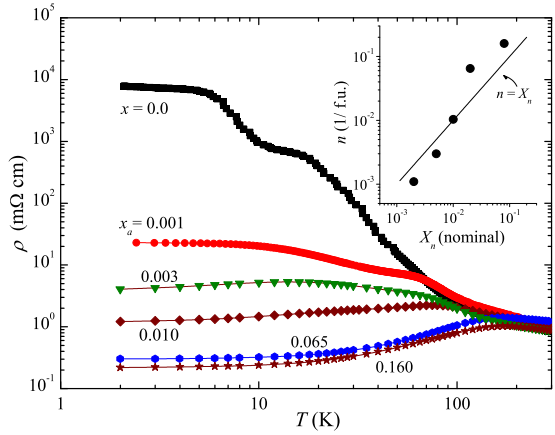}
\caption{\label{fig:epsart} Electrical resistivity $\rho(T)$ of FeSb$_{2-x}$Te$_x$ with varying actual doping concentration ($x_a$). Inset: correlation between carrier concentration $n$ and 
nominal Te content $x_{n}$, from which  $x_{a}$ is determined.} 
\end{figure}

The transition from a semiconducting to a metallic ground state at around $x_a$ = 0.003 is shown explicitly in Fig.\,1 by the evolution of $\rho(T)$ with $x_a$.  Upon 0.5\% ($x_a$ = 0.01) Te 
doping, $\rho(T)$ at below 10\,K is reduced by four orders of magnitude. Given the small energy gap of FeSb$_2$ ($\sim$6 meV), the SM transition to occur at an even smaller value of $x_a$ is 
not really surprising.  Above 150 K, by contrast, $\rho(T)$ is insensitive against doping. In this temperature range, the Arrhenius function holds true up to at least $x_a$ = 0.01 with a 
larger transport gap of 20$-$40 meV, equivalent to that of pure FeSb$_2$ \cite{bentien, sun_dalton}.  
The Hall coefficient $R_H$ (inset of Fig.\,2) is roughly constant at low temperatures for all the metallic samples from which, within a one-band model, the carrier concentration $n$ 
(=1/$e$$\mid$$R_H$$\mid$) can be estimated (inset of Fig.\,1). An anisotropic electrical transport in FeSb$_{2-x}$Te$_x$ was reported by Hu $et\,al.$ \cite{Hu_Te}. This however, is much 
weaker than the effect of the carrier concentration differing from sample to sample and will, therefore, be not addressed. For completeness, we list the directions of the applied 
heat/electrical current for all the samples employed: $\langle$001$\rangle$ (FeSb$_2$),  $\langle$110$\rangle$ (0.001), $\langle$001$\rangle$ (0.003), $\langle$110$\rangle$ (0.01), 
$\langle$100$\rangle$ (0.065) and $\langle$010$\rangle$ (0.160).

\begin{figure}
\includegraphics[width=0.9\linewidth]{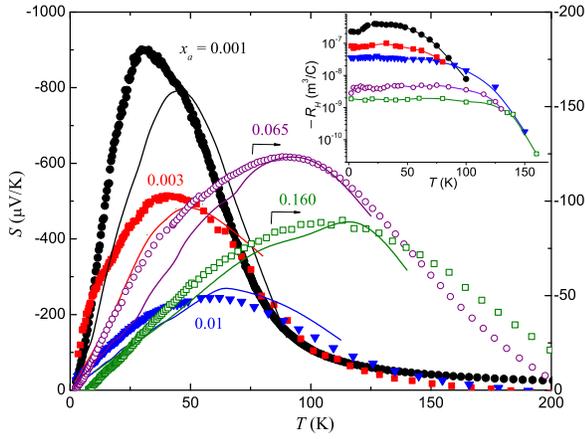}
\caption{\label{fig:epsart} Thermopower $S(T)$ of FeSb$_{2-x}$Te$_x$ with varying $x_a$. Solid lines are theoretical calculations based on the classical formula (see text) and further 
enhancement factors, i.e., 18, 22, 20, 28, and 32 for $x_a$ = 0.001, 0.003, 0.01, 0.065 and 0.160, respectively. Inset: Hall coefficient $-R_H(T)$ for FeSb$_{2-x}$Te$_x$ with symbols the 
same as in the main panel.  }
\end{figure}

As seen in Fig.\,2, the maximum absolute thermopower $|S_{max}|$ is diminished greatly upon increasing $x_a$. $|S_{max}|$ decreases from 6$-$45 mV/K at around 10 K in nominally pure FeSb$_2$ 
(refs. \cite{bentien, sun_prb, sun_dalton}) to 0.9 mV/K at 30\,K for $x_a$ = 0.001 and eventually to 90 $\mu$V/K at 110\,K for $x_a$ = 0.16, the metallic end member of this alloy series. 
Such a decrease of $|S_{max}|$  is not inconsistent with expectation, because the thermopower measures the energy derivative at the Fermi level of the electrical conductivity, which shows 
change by several orders of magnitude with Te doping (to a first order approximation,  $S$\,$\sim$\,ln$\rho$). 
However, here we put strong emphasis on one fact: by taking into account the respective carrier concentration $n$ determined from Hall coefficient (inset of Fig.\,2), the large enhancement 
of $S(T)$ as measured compared to classical expectations \cite{sun_prb}  still holds true when crossing the SM transition. In a parabolic band approximation, $S(T)$ of a degenerate electron 
system with dominant scattering by acoustic phonons is given by 
\begin{equation}
S(T)\,=\,\frac{\pi ^2}{3} \frac{k_{B}}{e} \frac{k_BT}{\epsilon _F},
\end{equation}
where the Fermi energy $\epsilon _F$ is a function of $n$ and $m^*$, 
\begin{equation}
\epsilon _F\,=\,\frac{h^2}{2m^*}(\frac{3n}{8\pi})^{2/3}.
\end{equation}
What is remarkable is that by assuming $m^*$ = $m_0$, the free electron mass, an almost universal factor 10$-$30 has to be multiplied on Eq. 1 in order to reproduce the experimental results 
of all the Te doped systems, as indicated by the solid-line curves (cf. Fig.\,2 and its caption). 

For simplicity, the above calculations were carried out up to only 100$-$150 K where $R_H$ is negative, whereas at higher temperatures multi-band corrections have to be considered. The 
enhancement to $S(T)$, on the other hand, can be regarded as corresponding reduction of $\epsilon _F$: for example in the case of $x_a$\,=\,0.065, $\epsilon _F$ estimated from Eq.\,1 (when 
applied to describe $S(T)$) is $\sim$180 K, a factor of 28 smaller than $\epsilon _F$\,=\,5000 K as estimated from the measured value of $n$ with the aid of Eq.\,2. 
Considerable deviations between the calculated lines and the measured $S(T)$ values are observed, particularly at temperatures below that of $|S_{max}|$. This is not too surprising, noting 
that the effects due to phonon drag, an additional impurity band as seen in FeSb$_2$ (ref. \cite{sun_prb}), as well as a possible nonparabolicity of the bands were ignored in the simple 
calculations. Instead of aiming for a good fitting to the experimental results, our emphasis is to understand whether and, if so, to what extent the aforedescribed thermopower is affected 
when crossing the SM transition. 

Our observations can provide valuable insight into the unusual TE transport of FeSb$_2$: the enhancement to thermopower found both in the semiconducting and metallic regimes most probably 
has a common microscopic origin, and should be considered an intrinsic electronic property.     
Arguments against a dominating phonon drag effect, in addition to the reasonable fitting of $S(T)$ as shown above, are the very different evolutions of $S(T)$ and $\kappa(T)$ (Fig.\,3) with 
$x_a$. While the values of both quantities are reduced, the maximum of the former shifts rapidly to higher temperature whereas that of the latter remains robust.  The factor 10$-$30 of the 
thermopower enhancement is too large to be explained by LDA band-structure calculations \cite{LDA}. A realistic scenario is to take into account a renormalization of $m^*$ by 
electron-electron correlations, similar to what occurs in heavy-fermion systems \cite{HF_tep}. This argument has found support from thermodynamic and magnetic measurements: for example, i) 
the electronic specific heat (not shown here) in combination with the measured carrier concentration yields an enhanced $m^*$ by a similar factor 10$-$30 for all the samples investigated; 
ii) the magnetic susceptibility above 100 K persists to be thermally activated when crossing the SM transition, indicating that the narrow gap and the correlated nature do not change.  

Given the importance of electron-electron correlations in the enhancement of $S(T)$, the underlying physics is yet to be clarified. Prevalent understanding is based either on local spin 
fluctuations as known for Kondo insulators or a nearly ferromagnetic semiconductor picture in an itinerant model \cite{petrovic1}. In particular for pure FeSb$_2$, where the charge carriers 
are assumed to be nondegenerate, our previous study \cite{sun_prb} revealed that its thermopower can be described by using not only the nondegenerate approximation, but also the degenerate 
approximation as given by Eq.\,1, with a similar enhancement factor. Of course, it remains to be shown whether the above interpretation of the enhanced thermopower based on a strongly 
renormalized $m^*$ can be applied to pure FeSb$_2$. Also, the observed maximum thermopower $|S_{max}|$ for $x_a$\,=\,0.001 and 0.003 exceeds 
$\frac{\pi^2}{3}\frac{k_B}{e}$\,$\approx$\,283\,$\mu$V/K, i.e., the classical upper bound expected from Eq.\,1 for degenerate electrons when $\epsilon _F$ $>$ $k_BT$, indicating that they 
might fall into the regime on the degenerate-nondegenerate borderline as well. Theoretical treatment of electron correlations in a nondegenerate approximation, however, is still an open 
issue \cite{Kotliar}.

\begin{figure}
\includegraphics[width=0.88\linewidth]{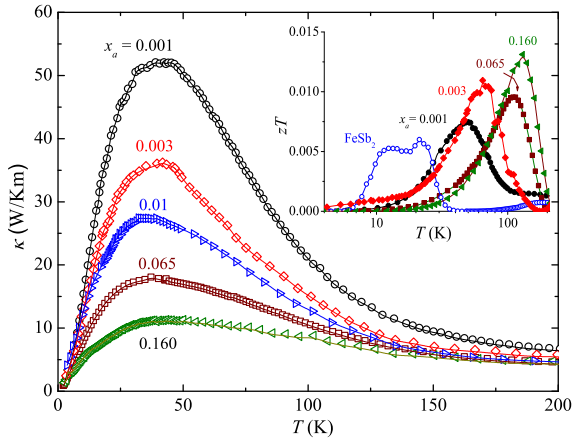}
\caption{\label{fig:epsart} Thermal conductivity $\kappa(T)$ of FeSb$_{2-x}$Te$_x$ with varying $x_a$. Solid lines indicate lattice contribution $\kappa _{L}(T)$ to each system. Inset: $zT$ 
vs $T$ for the different samples. }
\end{figure}

In view of the thermopower enhancement being almost unaffected upon crossing the SM transition, it is of great technological interest to see to which extent $\kappa$ can be reduced. As 
displayed in Fig. 3, there is a significant reduction of $\kappa (T)$ with increasing $x_a$. We attribute this reduction mainly to the introduced charge carriers rather than to chemical 
disorder: a small number of charge carriers in a pure semiconductor can dramatically reduce $\kappa$ as is known in, e.g., Ge \cite{Ge}. 
 What is certainly unfavorable for TE application is that, in the whole doping range, the lattice contribution $\kappa _L$ \cite{kappa} as indicated by lines in Fig. 3, is nearly equivalent 
to the measured $\kappa$. The electronic portion, $\kappa _e$, is less than 3\% of $\kappa$ at $T$$\leq$100 K for not only the semiconducting samples, but also the metallic ones. As a 
result, $zT$ shows only moderate increase, from 0.006 in FeSb$_2$ to 0.013 in $x_a$ = 0.160 (inset of Fig. 3), a value similar to that of the undoped FeSi \cite{mahan}. Practical TE 
materials are those with optimized $\kappa _L$ that is preferably less than or at least comparable to $\kappa _e$. 
As far as $\kappa _L$ dominates $\kappa$, as is currently observed, it turns out to be hard to realize a sufficiently large $zT$. 
Further nano-processing to reduce  $\kappa _L$ seems to be inevitable for possible application of this material. The preparation of thin film \cite{Ye1,Ye2} has already been initiated, 
however, well-controlled composition and purity have yet to be realized.  

To summarize, we have found that the large thermopower enhancement relative to the classical expectation previously observed for correlated FeSb$_2$ persists when crossing the SM transition 
induced by Te doping. In at least the metallic range, this phenomenon can be captured by the largely renormalized  charge carrier mass due to electron-electron correlations. 
Because of a dominant lattice thermal conductivity, $zT$ remains low in the FeSb$_{2-x}$Te$_x$ system. 
However, our study emphasizes the potential in exploring unconventional TE materials
through tuning a correlated system between semiconducting and metallic regimes. Along this line, searching for new correlated semiconductors with intrinsically low thermal conductivity 
(complex lattice structure, heavy atoms) is badly called for.


\begin{thebibliography}{99}
\bibitem{kappa} $\kappa _L$ ($=$ $\kappa$$-$$\kappa _e$) is estimated based on the Wiedemann-Franz law $\rho\kappa _{e}$/$T$ = $L_0$, with the Lorenz number $L_0$ = 2.44 $\times$ 10$^{-8}$ 
V$^2$/K$^2$.
\end{thebibliography}
\end{document}